# Topotactic fibrillogenesis of freeze-casted microridged collagen scaffolds for 3D cell culture

*Clément Rieu, Gervaise Mosser, Bernard Haye, Thibaud Coradin, Francisco M. Fernandes\*, Léa Trichet\**

Sorbonne Université, CNRS, Laboratoire de Chimie de la Matière Condensée de Paris, 4 place Jussieu, 75252 Paris Cedex 05, France

**Abstract**

Type I collagen is the main component of the extra-cellular matrix (ECM). *In vitro*, under a narrow window of physico-chemical conditions, type I collagen self-assembles to form complex supramolecular architectures reminiscent of those found in native ECM. Presently, a major challenge in collagen-based biomaterials is to couple the delicate collagen fibrillogenesis events with a controlled shaping process in non-denaturing conditions. In this work an ice-templating approach promoting the structuration of collagen into macroporous monoliths is used. Instead of common solvent removal procedures, a new topotactic conversion approach yielding self-assembled ordered fibrous materials is implemented. These collagen-only, non-cross-linked scaffolds exhibit uncommon mechanical properties in the wet state. With the help of the ice-patterned micro-ridge features, Normal Human Dermal Fibroblasts and C2C12 murine myoblasts successfully migrate and form highly-aligned populations within the resulting 3D biomimetic collagen scaffolds. These results open a new pathway to the development of new tissue engineering scaffolds ordered across various organization levels from the molecule to the macropore, and are of particular interest for biomedical applications where large scale 3D cell alignment is needed such as for muscular or nerve reconstruction.

\* Corresponding authors: lea.trichet@upmc.fr and francisco.fernandes@upmc.fr

# 1. Introduction

The fibrillar state of collagens is a key parameter for the mechanical properties of extracellular matrices and their interaction with cells. Being able to achieve fibrillogenesis in tissue-mimicking scaffolds is therefore of paramount importance to develop new materials for tissue engineering applications. *In vivo*, the synthesis and assembly of type I collagen into organized fibrillar structures is mediated by cells. In accordance with the local requirements of the tissue, collagen biosynthesis results in remarkably different supra-fibrillar architectures, ranging from highly-aligned domains in tendon to cholesteric arrangement of fibers in the bone. Reproducing these architectures *in vitro* requires a thorough control over the physico-chemical conditions before, during and after fibrillogenesis.[1] However, the assembly of collagen fibers into ordered domains in the micron range is in many cases insufficient as scaffolds also need a precise structuration at the macroscopic scale. To date, a vast range of techniques have been proposed to shape fibrillar collagen into dense materials, such as extrusion,[2] electrochemical alignment,[3] cell-mediated gel compaction[4] and plastic compression,[5] among others, with encouraging results already obtained *in vivo*. Still, these techniques often prove to be unsuitable for additional internal structuration of the materials. In particular porosity, a key morphological feature in biomaterials as it determines diffusion of nutrients and waste and can promote cell guidance, remains difficult to control, except by top-down 3D methods.[6] In this context, freeze-casting, a technique initially developed for the structuration of colloidal suspensions for the elaboration of ceramic materials,[7] has been adapted to produce scaffolds based on collagen and other biopolymers for tissue engineering purpose.[8–10] This technique enables to create oriented macroporous scaffolds from colloidal suspensions or polymer solutions by means of directional freezing followed by ice sublimation.[11] Upon anisotropic cooling, an oriented segregation occurs between the newly-formed ice and the solutes and/or particles initially present in the solution/suspension. Tuning of pore sizes and choice between isotropic or anisotropic global structures with anisotropic substructures can be achieved by playing with different parameters such as solution composition, freezing kinetics and temperature gradient.[12] For collagen-based scaffolds, static freeze-casting of type I collagen solutions with concentrations between 5 and 20 mg/mL[13–16] led to porosities ranging from 20 μm[16] to 200 μm at the upper extremity of the scaffold[15]. While these works have been successful in building porous materials out of acid soluble type I collagen, they have failed in inducing fibrillogenesis of the obtained porous collagen materials. As a result, the obtained scaffolds were water-soluble and required subsequent cross-linking, either by physical[13,17] or chemical[18,19] methods. The stiffness of the resulting materials usually ranges between a few kPa to a few hundreds of kPa in dry state but only between a few hundreds of Pa to a few kPa following rehydration.[20,21] As an alternative, fibrillogenesis of collagen can be induced before freezing. Using un-controlled freezing process in cylinders with a high aspect ratio, Lowe et al. demonstrated that comparable mechanical properties could be achieved in non-crosslinked fibrillar collagen aligned materials compared to crosslinked non-fibrillar scaffolds.[22] Fibrillogenesis prior to freezing was also used by Murray et al. in their successful clinical test to avoid risks associated with the use of chemical cross-linkers.[23–25] However, so far, no fibrillar macroporous collagen material presenting a controlled architecture at multiple length scales and of oriented porosity has been achieved.

In this work we have developed a new method to trigger type I collagen fibrillogenesis after freeze-casting, coupling the macroporous structure induced by the ice crystals to the self-assembly of the collagen molecules into fibers. In our topotactic process, ammonia vapors play a double role: while slowly melting ice crystals it induces fibril formation in the collagen domains condensed in between ice crystals. The obtained scaffolds present highly-organized suprafibrillar architecture and improved mechanical properties compared to non-fibrillar crosslinked collagen scaffolds reported in the literature.[20,21,26] These 3D scaffolds promote the proper alignment of C2C12 myoblasts and of Normal Human Dermal Fibroblasts *in vitro* and thus act as new 3D culture biomimetic environments with cell guiding properties.

## 2. Materials and Methods

**Collagen solution.** Collagen I was extracted from young rat's tail tendons. After thorough cleaning of tendons with PBS 1X and 4M NaCl, tendons were dissolved in 3mM HCl. Differential precipitation with 300mM NaCl and 600mM NaCl, followed by redissolution and dialysis in 3mM HCl achieved high collagen purity as confirmed by SDS PAGE. Final collagen concentration of 3.7±0.3 mg/mL was determined using hydroxyproline titration. Acid-soluble type I collagen stock solution was aliquoted and stored at 4°C. Prior to use, stock solution was centrifuged at 3000xg at 10°C in Vivaspin® tubes with a 300 kDa filter cut-off to reach a final concentration of 40 mg/mL.

**Freeze-casting**. Concentrated collagen solutions (0.2 mL, 40 mg/mL) were poured into cylindrical molds. After centrifugation each mold was individually freeze-cast on a home-made freeze-casting setup made of a copper rod partially immersed in liquid nitrogen, with a heating coil and a thermocouple to monitor the temperature of the cooled plate as previously described (Figure 1A).[27] The collagen solution was cooled at a -5°C.min$^{-1}$ rate from 20 to -60°C. The final temperature was kept at -60°C until the sample was entirely frozen. Samples were then kept in freezer at -80°C until further use.

**Collagen fibrillogenesis and sample preparation.** Fibrillogenesis of collagen was performed in different steps. First, the frozen samples were kept at 0°C under ammonia vapors for 48 h to ensure pre-fibrillogenesis induced by the pH increase. As ammonia vapors reach the ice/collagen surface, ice starts to melt thanks to the freeze point depression of the water/ammonia liquid mixture that is formed at the interface. Throughout the 48h contact time, where the sample is kept in a closed vessel of around 350mL with approx. 10 mL of 30 % ammonia solution, the ice front progressively recedes and the whole sample is exposed to ammonia. Samples (equivalent to approximately 0.2 ml initial volume) were then transferred to tubes floating on 2L water at 37°C and kept for 24h to remove ammonia. Samples were subsequently dipped into 40mL PBS 5X to complete fibrillogenesis at neutral pH for at least 2 weeks. The resulting scaffolds were 12 to 15 mm long with a diameter between 5 to 6 mm.

**Scanning electron microscopy.** Structural observations and quantitative analysis of the obtained structures were performed on dry samples lyophilized after freeze-casting. All the samples were coated with 10 nm gold layer before observation under the Hitachi S-3400N SEM (operating at 3 kV). Image analysis was performed using ImageJ software to determine the size of the pores on transverse sections cut across the length of the cylindrical scaffolds. Automatic image analysis was conducted after images were binarized to quantify the characteristic dimensions of the pores. Statistics on the length of major axis were obtained based on at least 600 different pores using Feret diameter measurement of individual pores (maximum caliper). The statistical description of the distance between the observed ridges was based on 20 individual measurements obtained from three different pictures of the two transverse sections considered. Whole scaffold longitudinal reconstitution was composed from 15 pictures taken at x27 magnification and superimposed using Adobe Photoshop.

The characterization of hydrated collagen samples after fibrillogenesis was conducted by cross-linking overnight in paraformaldehyde (PFA) 4 % vol. at 4°C, followed by 1h 2.5% glutaraldehyde in cacodylate buffer at 4°C. Samples were subsequently dehydrated using ethanol baths with increasing concentrations, and then super-critically dried.

**Transmission electron microscopy.** Hydrated samples were crosslinked with PFA, glutaraldehyde and osmium tetroxide 4 wt. %. They were subsequently dehydrated using baths with increasing concentrations in ethanol, progressively transferred to propylene oxide and incorporated in araldite resin prior to sectioning. Dry control samples obtained after freeze casting and lyophilization (no fibrillogenesis) were dipped in ethanol prior to the same protocol of inclusion with propylene oxide and araldite. After inclusion of samples in araldite, 500 μm thin sections were stained with toluidine blue and observed under a Nikon Eclipse E600Pol microscope equipped with a 2.5x, 10x and 40x objective and a DXM 1200 CCD camera as a first observation. Built-in cross-polarizers were used to observe the birefringency of thin sections. Then, 70 nm ultrathin sections (Leica microtome) were contrasted with uranyl acetate and observed on a transmission electron microscope FEI Tecnai Spirit G2 operating at 120 kV to observe the ultrastructural collagen features. Images were recorded on a Gatan CCD camera.

**Mechanical testing.** Tensile testing was done on an Instron 3343 machine equipped with a 10N load-cell and a watertight column to perform the tests in hydrated conditions. The traction was exerted on rectangular sections cut from the center of the samples by means of a vibration microtome (HM 650V Thermo Scientific), with a blade vibrating at 100Hz and with a cutting speed of 2mm/min. The 1.5 mm slices obtained were kept in PBS to ensure equilibration with the solution. The tests were performed at a velocity of 1 mm.min$^{-1}$ along the freezing direction. The samples were held vertically in PBS solution between Velcro straps glued to the clamping jaws of the machine holders. The initial cross-section of each slice was precisely measured by taking a picture of this cross-section on a calibrated contact angle machine (Krüss, DSA30) and measuring the surface using ImageJ software.

**Cell culture.** Cell culture was performed on whole samples cut at each end to remove the characteristic irregular porosity zone due to supercooling during freeze casting. Samples were first rinsed several times for at least 24 hours in cell culture medium. Cells were then seeded on top of the scaffolds lying horizontally in the medium in a 12-well culture plate, which allowed the investigation of cell penetration and migration inside the scaffolds. Normal Human Dermal Fibroblasts (PromoCell™) at passage 13 were seeded, with 110 000 cells per well, and left 3 weeks in culture, changing medium (low glucose Dulbecco's modified Eagle's medium DMEM, GIBCO, 10% fetal bovine serum, GIBCO) every other day. Longitudinal slices of the scaffolds were also seeded in the same time and conditions. C2C12 cells (ATCC) were seeded at P13 with 66 000 cells per well. Two conditions were used. Half were left in proliferation medium (high glucose DMEM, GIBCO, 20% fetal bovine serum, GIBCO) for 11 days in culture while the other half was cultivated during the same period of time but with differentiation medium (high glucose DMEM, GIBCO, 2% horse serum, GIBCO) for the last four days. All cell culture conditions were carried out in triplicate.

**Fluorescence and second harmonic generation (SHG) microscopy.** At the end of the culture, cells were rinsed with PBS1X and left overnight in 4% PFA to ensure complete cross-linking of the whole scaffold. 1.2 mm slices were then cut from the fixed samples using a vibration microtome. For both types of cells, fluorescent labeling of the nuclei (DAPI, Invitrogen) and the actin filaments (Alexa Fluor® 488 phalloidin, Invitrogen) was performed. Additional labeling of C2C12 with MF-20 hybridoma mouse IgG2B primary antibody and Alexa Fluor® 546 goat anti-mouse IgG2B secondary antibody (Invitrogen) was used to investigate their differentiation.

Extensive observations of the samples were carried out under fluorescence microscope (Axio Imager D.1, Zeiss). Further observations were conducted on a Leica SP5 upright Confocal, multiphoton laser scanning microscope which enabled to combine fluorescence and second harmonic generation (SHG) observations.

## 3. Results

### 3.1 Ice-templating of macroporous collagen scaffolds

The home-built freeze casting set-up enabled to control the growth of ice crystals in 40 mg.mL-1 type I collagen solutions and thus to successfully template the resulting scaffold. After lyophilization, a porous scaffold with lamellar pores oriented along the freezing direction and presenting a gradient of porosity was obtained (Figure 1B). At the bottom of the scaffold, pores were arranged in small domains of parallel sheets 25 to 95 µm wide ($10^{th}$-$90^{th}$ percentile) (Figure 1B bottom right, Figure 1C orange), randomly distributed in 2D (Supporting Information Figure S1A,C). Following ice growth direction, the domains of parallel sheets broadened (Supporting Information Figure S1B,E) and pore major axis increased between 50 and 270 µm ($10^{th}$-$90^{th}$ percentile), with some pores reaching up to 1.3 mm (Figure 1B top right, Figure 1C violet). A substructure made of parallel ridges also ran along the same direction of the scaffold (Figure 1D,E). The ridges only appeared on one side of a wall and were periodically spaced, with the period ranging from 6.9 to 10.0 µm ($10^{th}$-$90^{th}$ percentile), at the bottom, to 14.5 to 18.9 µm ($10^{th}$-$90^{th}$ percentile) at the top of the scaffold (Figure 1F).

### 3.2 Topotactic fibrillogenesis of ice-templated collagen scaffolds

After ice-templating (Figure 2 step 1), the samples were kept for 48 h under ammonia vapor at 0°C, turning the frozen scaffold into a translucent hydrogel (Figure 2 step 2). After being placed under a water-saturated atmosphere at 37°C for 24 h (Figure 2 step 3), the hydrogel remained translucent and turned opaque upon further dipping into PBS 5X (Figure 2 step 4). TEM images of samples after freeze-casting showed liquid crystal-like local organization of collagen as well as the absence of fibrils (Figure 2 bottom left). After ammonia treatment, small fibrils were visible displaying some degree of local organization, reminiscent of the previously-described liquid crystal organizations of collagen[28–30] (Figure 2 bottom middle). At the end of the process, after steps 3 and 4, the previously-observed dense organized areas of small fibrils (Supporting Information Figure S2A,B) coexist with domains comprising larger fibrils (Figure 2 bottom right, Supporting Information Figure S2C). Pores remained open after this process (Supporting Information Figure S3A) as opposed to dipping freeze-casted scaffold directly after step 1 into a neutral buffer such as PBS 10X that resulted in occluded pores (Supporting Information Figure S3B).

### 3.3 Scaffold morphology and mechanics

The opaque water-stable self-standing scaffolds obtained at the end of the above-described process were 12 to 15 mm long and 5 to 6 mm wide (Figure 3A). Second Harmonic Generation (SHG) imaging of slices of scaffolds gave strong signal characteristic of aligned and/or fibrillar collagen. SHG enabled the 3D reconstitution of the aligned porous structure of the scaffolds (Figure 3B). Scanning electron microscopy (Figure 3C,D) confirmed these observations by showing open pores with ridges similar to the ones observed on freeze-cast scaffolds, evidencing the topotactic nature of the process. In addition some collagen fibers could be visualized inside the pores in certain regions of the scaffold. Observation of thin sections under cross-polarizers showed birefringency of the scaffold ridges, with a maximum transmission at the same angle for all the ridges (Figure 4A,B and Supporting Information Figure S3A, for a comparison without polarizer). On TEM pictures, scaffold walls present different structural domains, with tightly packed and aligned small fibrils (Figure 4C, Supporting Information Figure S2A) and less dense domains with larger fibrils that display wave patterns (Figure 4D, Supporting Information Figure S2C).

Fibers outside the wall present the characteristic 67 nm D banding (Figure 4D, Supporting Information Figure S2B).

Mechanical properties were measured performing tensile tests in PBS 1X on longitudinal slices from different samples. Figure 5A displays a typical stress vs. strain curve with corresponding photos of the sample in Figure 5B. The sample undergoes a linear deformation (Figure 5B1,2) to approximately 105±28 % strain and 33±6 kPa stress in average (Table 1) with a Young modulus reaching 33±12 kPa. After this point, it finally cracks at several levels provoking the delamination of the sample (Figure 5B3,4).

**Table 1.** Mean value of the Young modulus, strength and strain at failure from the different tests, and deviation (twice the standard deviation).

|  | Number of tests | Mean value | Deviation ($2\sigma$) |
|---|---|---|---|
| Young modulus (kPa) | N=5 | 33 | 12 |
| Ultimate Tensile Strength (kPa) | N=4 | 33 | 6 |
| Strain at failure (%) | N=4 | 105 | 28 |

**3.4 *In vitro* 3D cell culture**

*3.4.1. Normal Human Dermal Fibroblasts*

Normal Human Dermal Fibroblasts were seeded and cultured on entire scaffolds to observe cell colonization (Figure 6A). After three weeks in culture, they showed high proliferation on the scaffold surface, forming a dense layer of cells. Most cells seemed aligned along the principal direction (direction of freezing) as shown on Figure 6A1. When observing the outside of a thin longitudinal upper slice, cells are aligned in the same direction as the ridges (Figure 6A2). This is confirmed by confocal microscopy coupled with SHG, as seen on Figure 6A3 and on Supporting Information Movie S1A, where cells, which actin filaments are labeled, all adopt a fusiform shape aligned along the main direction of the scaffold. Recovering internal longitudinal slices of the cellularized scaffolds allowed for the study of deeper colonization. A first and most obvious cell population originated from cell migration through the open pores at each extremity of the scaffold (Figure 6A4 top left). Cells migrated inside the scaffold over a distance up to 1.5 mm, they aligned along the main direction (Figure 6A5) and adopted fusiform shapes with long cytoplasmic protrusions (Figure 6A6). It was observed that a deeper colonization involving a higher number of cells occurred from the extremity with the larger pores (top of the scaffold, as displayed on Figure 1B) compared to the one with the smaller pores (bottom of the scaffold). A second type of colonization also occurred through the sides of the scaffold, resulting in a density gradient of cells toward the inner part of the scaffold (Figure 6A4 bottom) over a distance up to 500 µm. In this region, fibroblasts aligned along the main direction as shown on Figure 6A7 and on Supporting Information Movie S1B. Confocal microscopy enabled to observe cells in the plane of a wall and to show their alignment along the ridges (Figure 6A8, Supporting Information Movie S1C). Noticeably, when similar NHDF cultures were performed on pre-cut longitudinal slices, high proliferation and significant cell alignment were also observed (Figure 6B1,2).

*3.4.2. C2C12 myoblasts*

Similar colonization experiments were carried out using C2C12 murine myoblast cell line. Cells were seeded on whole scaffolds for 11 days. Half of the samples were left in proliferation medium whereas the other half was supplemented with differentiation medium for the last 4 days instead of high serum concentration proliferation medium. No significant differences could be observed between the two conditions. In each case, high proliferation and alignment of myoblasts was observed outside the scaffold (Figure 7A1,2), in the same way as with NHDF. Some parallel myotubes expressing heavy myosin chains (MF20) could sometimes be observed as indicated by immuno-staining (Figure 7A). Same internal colonization patterns as for NHDF could also be observed from both scaffold extremities and sides but over a longer distance, reaching 2.5 mm from the extremities and 1.5 mm from sides compared to 1.5 mm and 500 μm for NHDF, respectively (Figure 7B1,2). Some myotubes positive for MF20 were observed aligned to the main direction of the scaffold (Figure 7C1,2). Confocal microscopy reconstitutions as shown on Figure 7D1,2 and Supporting Information Movie S2A and Movie S2B illustrate both types of colonization: from an extremity (Figure 7D1 and Supporting Information Movie S2A), and from a side (Figure 7D2 and Supporting Information Movie S2B). They also enable to observe the anisotropic shape of the myoblasts.

## 4. Discussion

### 4.1 Scaffold structuration and properties

To address the current challenge of preparing scaffolds combining fibrillar collagen and oriented porosity, we have developed a new topotactic procedure to induce collagen fibrillogenesis after freeze casting and obtain scaffolds presenting anisotropic features at multiple length scales. The protocol enables to produce stable scaffolds in physiological conditions without crosslinking or dehydration process and therefore appears relevant to mimic collagen structures observed in native tissues while preserving collagen bioactivity.

A first challenge addressed was to induce the self-organization of type I collagen molecules into liquid crystal phases within the frozen materials. In collagen solutions at 3 mM HCl, isotropic/anisotropic phase transition was reported to occur at *ca*. 50 mg/mL.[28] Here, while the initial collagen concentration is smaller (40 mg/mL), phase segregation due to ice crystallization locally concentrates collagen above the critical value, as confirmed by the observation of liquid crystal-like organizations after ice templating. In parallel, the ice crystal growth generated ridges parallel to the main axis. These are the first reported ridges in collagen-based scaffolds obtained by directional freezing. Such features have been previously observed in porous ceramics, where they might even form bridges between lamellae.[7] Micro-ridged pores were also observed after freeze-casting chitosan solutions, with a regular spacing of approximately 10 μm.[9] These overgrown dendrites supposedly arise from local ice crystal tip splitting, due to particle-interface interactions occurring in dense solutions. In fact, this phenomenon is highly dependent on the solvent nature and was not observed when using collagen solutions prepared in acetic acid (data not shown).

Our goal was to stabilize the sample with its hierarchical organization *via* collagen fibrillogenesis using a topotactic process, *i.e.* without changing the structure and sub-structure of the walls and pores. Here, the term topotaxis strictly refers to the above definition used in the field of chemistry and not to the single-cell sensitivity to the topography gradient phenomenon introduced by Park *et al.*,[31] which was not addressed in the present work. Dipping the freeze-casted scaffold directly into a neutral buffer such as PBS 10X resulted into a loose material with occluded pores, indicating an unfavorable competition between collagen fibrillogenesis and dissolution (Supporting Information Figure S3B). Similar results were obtained when scaffolds were lyophilized before being placed in the buffer medium. To circumvent this issue we turned our attention to a gas phase process using ammonia vapor. This strategy has been widely used to induce fibrillogenesis in concentrated type I collagen solutions at room temperature.[32,33] In our case, using ammonia vapors presents another key advantage: it lowers the freezing temperature of ice. This freezing point depression allows to progressively achieve fibrillogenesis of the collagen within the walls of the sample kept at 0 °C (Figure 2 step 2). After 48 h under ammonia vapor, the recovered hydrogel was self-standing and translucent, suggesting that small collagen fibrils had been formed. TEM images confirmed that liquid-crystal organization was preserved during fibrillogenesis. This assumption was supported by observation of semi-thin sections under polarized microscopes showing birefringence. Two specific patterns stand out from the observations made: the alignment of fibrils on large domains which can reach dimension above the tens of micrometers (Figure 4C and Supporting Information Figure S2A), and the presence of wave or arch pattern (Figure 2, Figure 4D and Supporting Information Figure S2C). The first one would correspond to the stabilization of nematic phases while the second pattern would correspond to cholesteric or precholesteric phases.[29] Both phases have been previously observed in highly concentrated collagen solutions as well as in the stabilized fibrillary state.[28,29] Comparison with the literature suggests that the high compaction of the network together with the small size of fibrils would correspond to collagen concentration higher than 300 mg/mL.[30]

The following water vapor step used to rinse off $NH_3$ and decrease the pH of the aqueous phase turned out to be paramount to preserve an open porosity. Dipping the scaffold into PBS 5X directly after exposure to ammonia vapors (step 2) resulted in a swollen translucent scaffold with occluded pores. This may be due to the fact that, since the pH of the solution remains high, collagen molecules and fibrils still bear a significant negative charge and can be destabilized by a rapid increase in ionic strength. In contrast, after the water vapor step has allowed decrease ammonia content, the scaffold can be dipped into PBS 5x without significant change in pH and thus without destabilization. This strategy further favors the growth of fibrils, as evidenced on the TEM images. Noticeably, the macroscopic aspect change of the scaffold from translucent to opaque started instantaneously and could be followed by naked eye observation for a few hours. Nevertheless, the scaffold was left in the buffer medium for 2 weeks before use to ensure complete stabilized fibrillogenesis of the scaffolds as collagen matrix can remodel for days after neutralization.[34,35]

From a morphological standpoint, the macroporous materials presented here could be compared to collagen sponges. However, collagen sponge lack the fibrillary structure attained here. Nevertheless collagen sponges are still the most common form for collagen biomaterials used in clinical applications, such as the artificial skin graft Integra®.[36] These materials are usually prepared by freezing a solution of collagen and removing water by lyophilisation or by using an organic solvent.[37] In the absence of collagen fibrillogenesis, the resulting sponges are mechanically weak and soluble in water, requiring additional cross-linking steps [38]. Resulting materials have typical Young moduli of 10 kPa in the dry state that systematically decrease in the wet state.[21,39] While mechanical properties can be improved by addition of glycosaminoglycan (GAG) and additional processing steps,[20,26] the Young modulus value of 33 kPa obtained here is unprecedented for pure collagen hydrated macroporous scaffolds. This emphasizes the benefits of our described protocol that allows for the preparation of chemically-stable and mechanically-strong collagen materials without any required additional cross-linking step. Moreover, these scaffolds can be easily optimized in terms of pore size and collagen concentration to tune their mechanical properties according to the targeted colonizing cells.

**4.2 Guiding properties for the cells**

The high anisotropy of the pores and the ridges substructure, parallel to the direction of freezing, makes the obtained scaffolds appealing to promote cell alignment, a feature highly sought after in tissue engineering, in particular for cells such as myoblasts, tenocytes or neurons.

In our experiments we observed an alignment of both C2C12 and NHDF cells along the main direction of the scaffold inside and outside the scaffolds. Cell alignment outside the scaffold may come from the outer grooves observed under SEM (Supporting Information Figure S4). Inside, cells colonize from both the sample extremities and the sample sides. In both cases cells align along the freezing direction. While the main migration route for cell colonization is considered to be the materials' macroporosity, it should also be possible for NHDF to migrate through the scaffold walls *via* collagen degradation/remodeling.[40]

The partial colonization of scaffolds internal space without any physical constraints nor biochemical cues except the scaffold's oriented porosity itself to guide the cells in is highly encouraging. While migration from both ends of the material was evidenced, it was not possible to observe fully colonized channels. This may be explained by the kinetics of colonization: here, a migration of NHDFs over *ca*. 500 µm was obtained after 11 days of culture, and C2C12 myoblasts could colonize the scaffolds over a three times larger distance. Comparable data for fibroblast or myoblast migration in 3D environments are scarce in the literature because colonization rate is expected to depend on many parameters including exact matrix composition and stiffness.[41] However, NHDFs seeded in diluted collagen hydrogels (2 mg.mL$^{-1}$) within microfluidic devices were reported to migrate at rates between 20 and 200 µm per day.[42,43] While our own measured rate (*ca*. 50 µm per day) lies in this range, our system differs in the initial cell seeding conditions (i.e. pre-immobilization vs. adhesion), the absence of flow as well as the size and orientation of the pores. Still, questions may be raised about the possibility to colonize the whole scaffold due to a possible lack of diffusion of nutrients, factors and wastes in the core of the scaffold in spite of the high porosity. Longer culture time and thorough follow-up of colonization over time may answer these questions. The colonization of our scaffolds might also be further improved by the adjunction of signaling biomolecules.[44]

The effect of scaffold structuration on cell behavior has been previously studied for many different types of cells. Caliari *et al.* showed that anisotropic collagen-glycosaminoglycan scaffolds induced higher proliferation rate and metabolic activity of equine tendon cells, as well as better alignment, compared to isotropic counterparts.[45] The size of the pores also had an effect on cell activity, in particular regarding cell entry into the scaffolds. The smallest pores, around 55 µm, induced limited cell penetration as compared to 152 and 243 µm ones. This is in agreement with our observations of a higher number of cells and deeper colonization through the large pores (50 to 270 µm wide) compared to the small ones (25 to 95 µm). Gonnerman *et al.* have shown that cardiomyocyte alignment and beating activity was enhanced on larger pore size constructs.[46] However, tenocyte metabolic activity was reported as being more efficient with smaller pore size.[47] Thus the optimal pore size depends on cell type and freeze casting should offer the possibility to fine-tune it to the desired dimensions.

The generation of micro-ridges regularly spaced on the scaffold walls and oriented parallel to the pore main axis appears to confer an additional advantage to our process as it favors cell alignment by topographical contact guidance. A similar observation was reported for freeze-casted chitosan scaffolds which strongly favored the alignment of chick root dorsal neurites.[9] A particularly striking result we obtained is the clear alignment of NHDFs cells seeded on scaffold slices, *i.e.* in pseudo 2D conditions. Indeed, surface patterning either on coated PDMS[48,49] or pHEMA[50] substrates, or from microgrooved collagen scaffolds[51] has been shown to increase muscular cell alignment and the formation of myotubes, with the possibility to optimize the spatial periodicity of the features.[48] Here the secondary guidance provided by the ridges is intrinsic to our materials and can also be optimized depending on the targeted cells.

The oriented porous collagen scaffolds presented in our study could also open the road to new strategies for the *in vitro* development of larger skeletal muscle 3D tissues with aligned myofibers. The production of such artificial tissues relies on cell mediated and guided gel compaction,[52–54] and on the transfer of aligned cell sheets into hydrogels,[55–57] among other techniques. In our study the ability of C2C12 cells to differentiate and merge into myotubes was evidenced with limited success. However, it is likely that a longer differentiation time would allow the formation of myotubes, particularly inside the scaffold where diffusion may delay the change of environment and thus differentiation. Still, this may not fully account for the low differentiation and the main hypothesis is the lack of other ECM components, such as laminin, that were shown to play a paramount role in muscle development.[58] The here-presented scaffolds could then be improved with the addition of these other extracellular matrix proteins to the initial collagen solution or by coating them on the preformed scaffold. [9,59] Pore size and collagen concentration might also direct the mechanical properties of myoblasts' microenvironment and hence their differentiation ability, which was shown to be optimal on gels with stiffness around 12 kPa.[60]

5. Conclusion

We have described here a processing strategy that enabled to obtain highly porous and anisotropic fibrillar collagen biomaterials, with pores patterned with regularly-spaced ridges running all along the scaffold, *via* a topotactic transformation. The self-standing scaffolds exhibit unprecedented stiffness for macroporous pure-collagen scaffolds in the wet state, sparing the use of any cross-linking technique. They showed high bioactivity towards NHDF and C2C12 cells that can colonize these scaffolds without additional physical, chemical or biological cues. The demonstrated ability to align cells could be of significant interest for muscle, tendon or nerve regeneration. In particular, the pore size gradients generated by the freeze-casting process could be used for the colonization and controlled differentiation of different types of cells and the generation of junctions between different tissues.

**Acknowledgements**

The authors would like to thank Costantino Creton, Jingwen Zhao, and Francisco Cedano for their help with the tensile testing experiments. Heng Lu, Yupeng Shi and Christophe Hélary are kindly acknowledged for their help with cell culture. The authors also thank Jennifer Coridon and Julien Dumont for help with sample preparation using vibratome and image acquisitions using confocal fluorescence and SHG microscopy. C. R. was supported by a French Ministère de la Recherche fellowship.

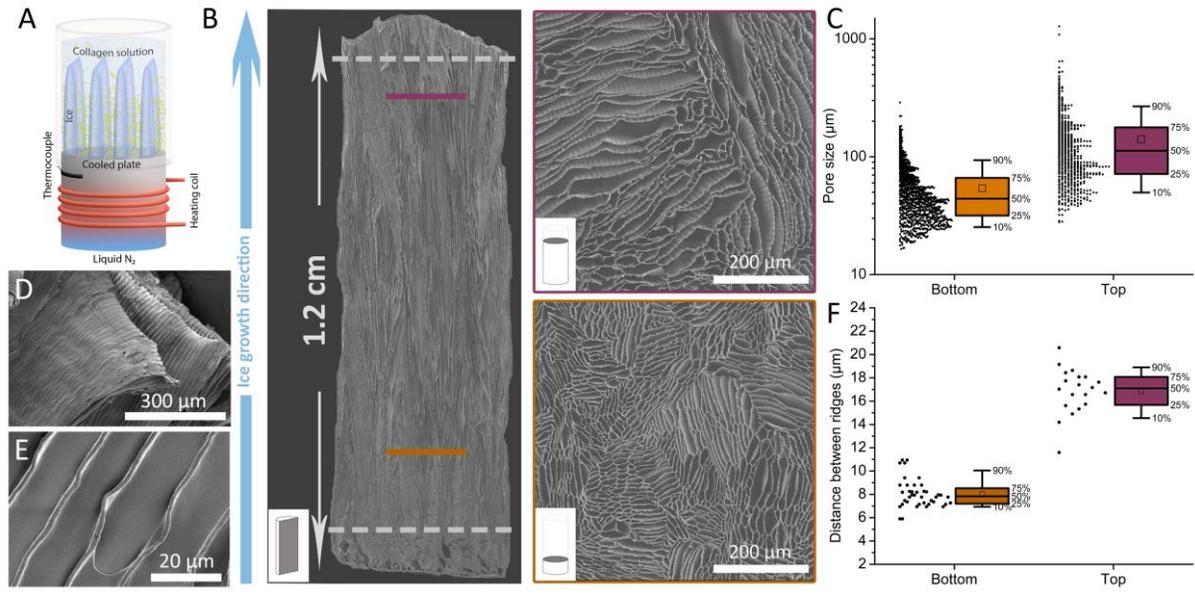

**Figure 1.** Collagen freeze-casting and resulting structures of the lyophilized scaffold. (A) Schematic of the setup enabling the growth of the ice crystals and the subsequent concentration of collagen in walls due to phase segregation. (B) SEM reconstitution from multiple images of the longitudinal section of the scaffold resulting from freeze-casting, observed after lyophilisation (left) and two transverse sections at different heights showing different pore sizes (right). Central part delineated by the dashed lines was used for cell culture. (C): Pore size quantification by Feret maximum diameter measurement of the two cross-sections displayed on (B). (D)(E) SEM images of the pore walls patterned with regular parallel ridges running all along the main axis of the scaffold, observed at two different magnifications. (F) Quantification of the distance between ridges in the two different sections displayed in (B).

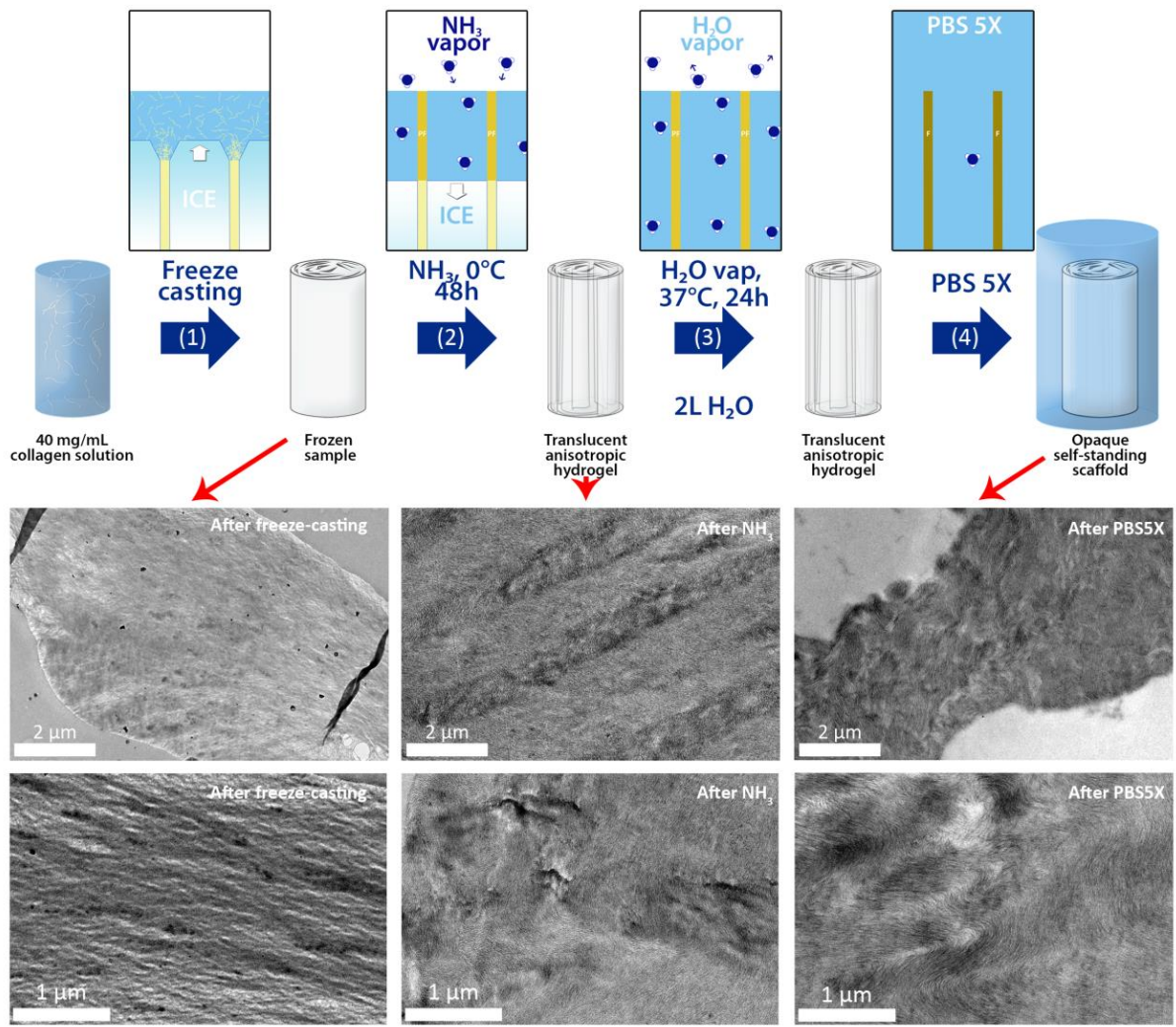

**Figure 2.** Process from freeze-casted sample to fibrillar scaffold. Center: schematic of the scaffold after each step from solution to the fibrillar scaffold. Top: schematic of the transformation occurring during each step. The collagen (yellow) solution is first ice-templated (1) then kept at 0°C for 48 hours under ammonia saturated atmosphere to melt the ice and trigger pre-fibrillogenesis of the collagen (2). Ammonia is then desorbed from the sample by placing it in an open vessel floating on 2L of water at 37°C (3) and the scaffold is finally placed into PBS 5X to ensure its complete fibrillogenesis. Bottom: TEM observations of the scaffold after steps (1), (2) and (4) respectively, and at two different magnifications.

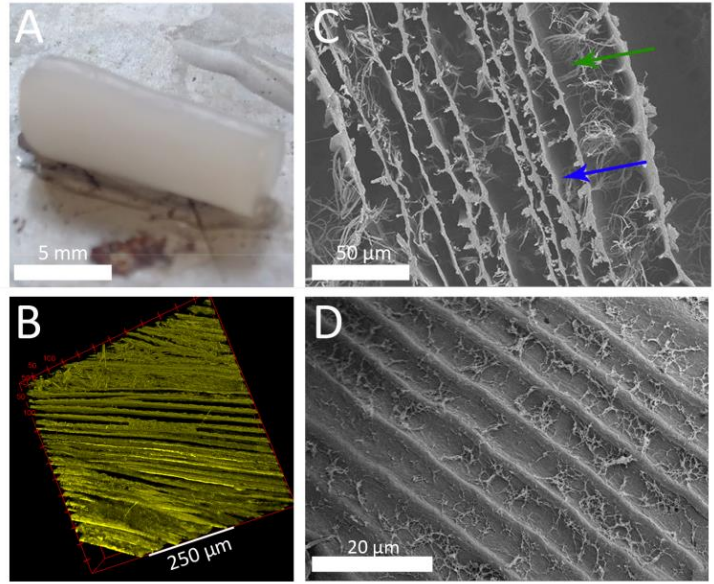

**Figure 3.** Resulting self-standing scaffolds presenting fibrillar collagen with preserved pore and micro-ridge structures. (A) Picture of the opaque self-standing scaffold after the fibrillogenesis protocol. (B) 3D spatial reconstitution of a slice of scaffold observed under confocal second harmonic generation (SHG). (C) (D) Scanning Electron Microscopy pictures of a slice of the scaffold (blue arrow: a ridge, green arrow: a self-standing fiber).

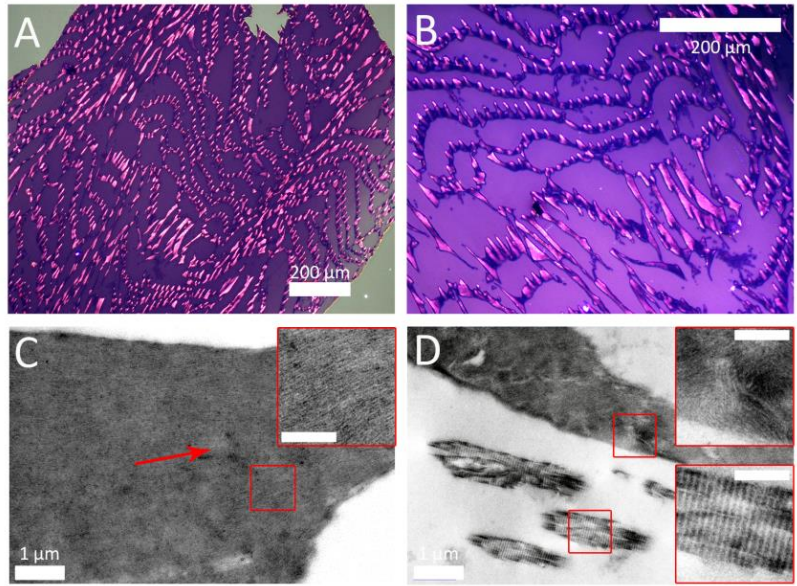

**Figure 4:** Resulting micro-structure of the fibrillar scaffold. (A) (B) Images of semi-thin sections of the scaffold observed under crossed polarizers. (C) (D) Transmission Electron Microscopy images of a section of the scaffold displaying different sizes and densities of fibers (red arrow: direction of the fibers on the picture). Insets: scale bar = 200 nm.

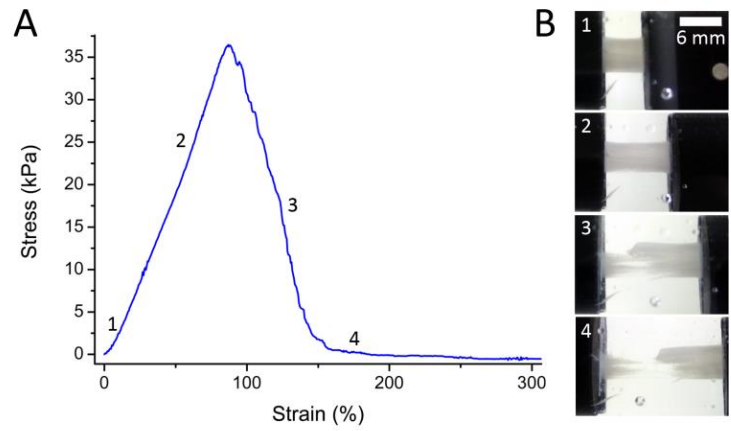

**Figure 5.** Mechanical properties of the scaffold after fibrillogenesis. (A) Stress vs. strain curve of a 1.5mm thick slice of scaffold pulled at a velocity of 1 mm/min. (B) Pictures of the scaffold corresponding to the numbers indicated on the curve.

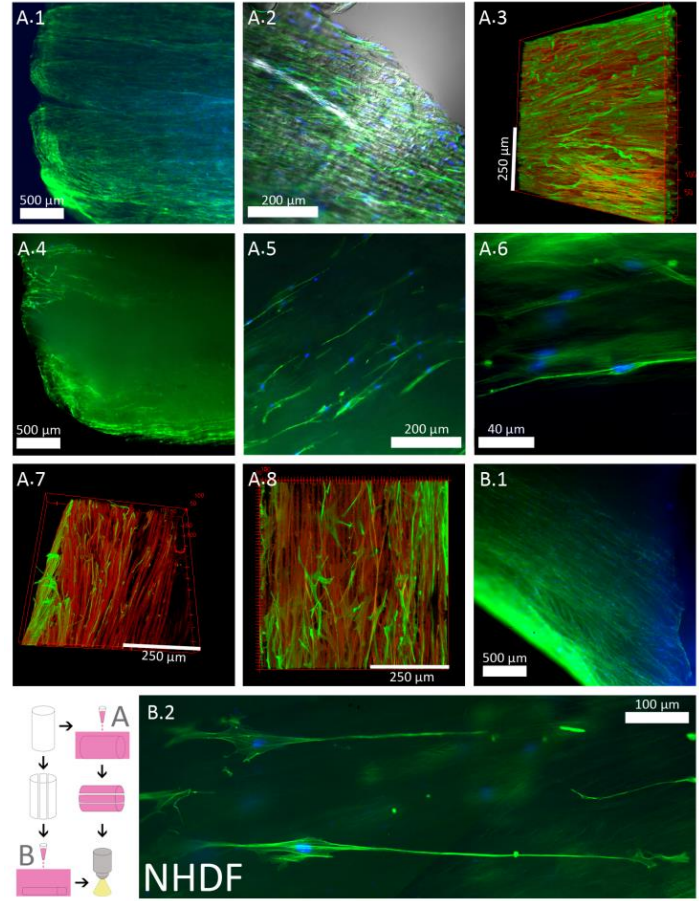

**Figure 6.** Normal Human Dermal Fibroblasts (NHDF) colonization of whole scaffold (A) and central pre-cut slice (B). (A1) Fluorescence microscopy image of NHDF proliferation on the outside of the scaffold. (A2) Same condition at higher magnification with the brightfield channel. (A3) SHG and fluorescence 3D reconstitution from confocal microscopy of the outer part of the scaffold (longitudinal upper slice) (corresponds to Supporting Information Movie S1A). (A4-6) Fluorescence images of NHDFs inside the scaffolds at different magnifications (central slice). (A7) SHG and fluorescence 3D reconstitution from confocal microscopy of the side of the scaffold with a cell density gradient (corresponds to Supporting Information Movie S1B). (A8) Observation of cells and ridges on a wall (corresponds to Supporting Information Movie S1C). (B1) Fluorescence image of a central slice of scaffold cut prior to seeding. (B2) High magnification of the same sample. Green: Actin; Blue: DAPI; Red: collagen SHG signal. Visualization window for 3D reconstitutions: 591 x 591 μm.

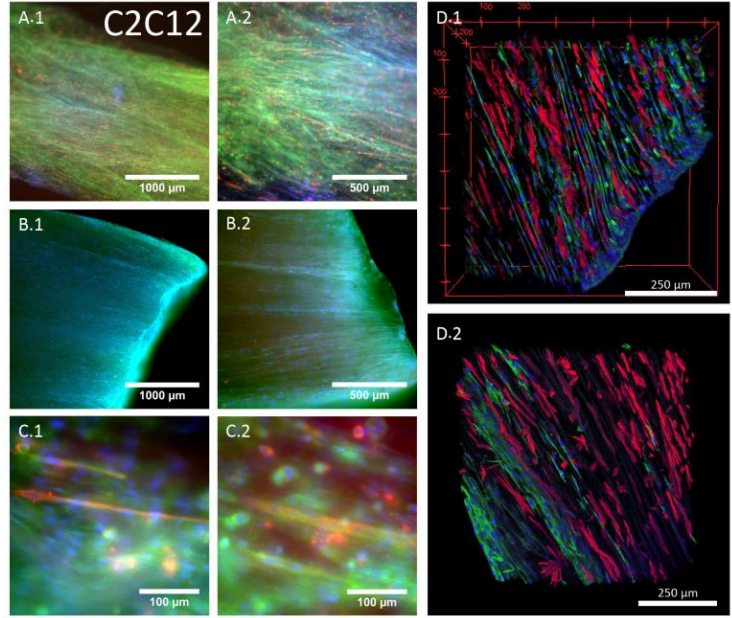

**Figure 7.** C2C12 myoblast colonization of whole scaffold. (A1) (A2) Fluorescence microscopy images of C2C12 proliferation on the outside of the scaffold at two different magnifications. (B1) (B2) Same technique on the inside of the scaffold at two different magnifications (longitudinal central slice). (C1) High magnification images of myotube formation (longitudinal outer slice). (C2) High magnification images of myotube formation (longitudinal central slice). (D1) Confocal SHG/fluorescence 3D reconstitution of C2C12 colonization inside the scaffold from one of its extremities (longitudinal central slice) (corresponds to Supporting Information Movie S2A). (D2) Same technique used on a scaffold side (corresponds to Supporting Information Movie S2B). For fluorescence 2D images: green: Actin; blue: DAPI; red: MF20. For 3D reconstitutions: green: actin; blue: DAPI; red: SHG signal. Visualization window for 3D reconstitution: 738x738 μm.

**Supporting Information**

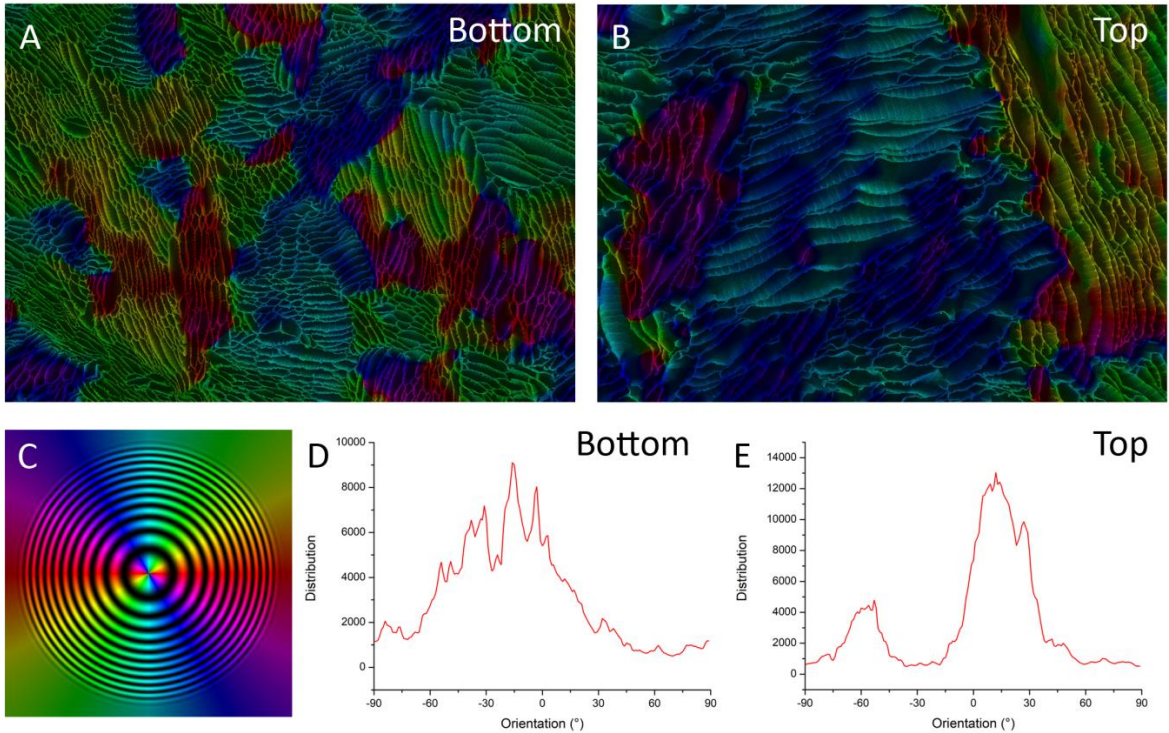

**Figure S1.** Orientation of pore domains at two different positions of the scaffold. (A) Pore domain orientation on the SEM image of the cross-section at the bottom of the sample, displayed in Fig.1.B, analyzed and colored using the orientation analysis plugin (OrientationJ) in ImageJ software v1.51n.[S1] The orientation mapping was calculated using a 20 pixel Gaussian window and the spatial derivatives were calculated using a cubic spline interpolation. (B) Same analysis performed on the image of the cross-section at the top of the scaffold (Figure 1B). (C) Angular color panel used to color pore domains in function of the normal vector orientation. (D) Angular distribution of the pore domains after analysis with ImageJ of the bottom section in (A). (E) Angular distribution of the pore domain after the same analysis of the top section in (B).

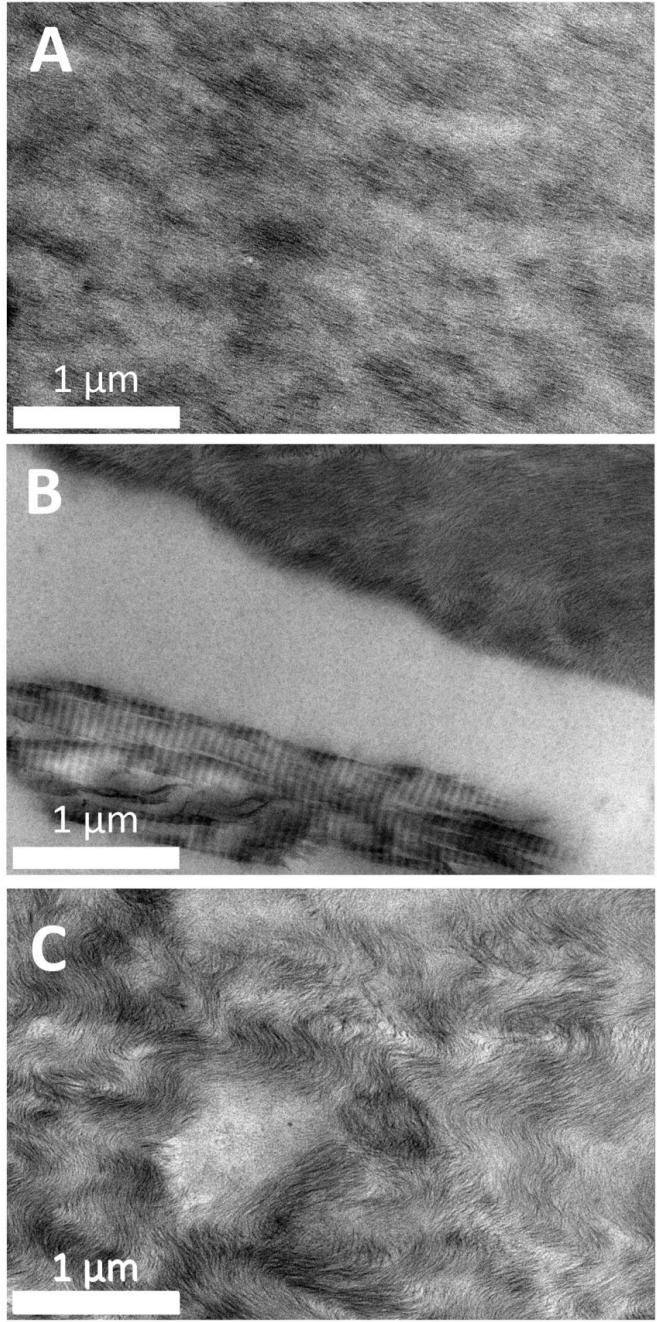

**Figure S2.** TEM pictures of the different structures observed in final scaffolds in PBS 5X (after step 4 as displayed on Figure 2). (A) Picture of a unidirectional alignment of collagen fibrils. (B) Picture of a wall with the same unidirectional alignment (top) and a striated fiber in between walls. (C) Picture of arch-like patterns formed by fibrils.

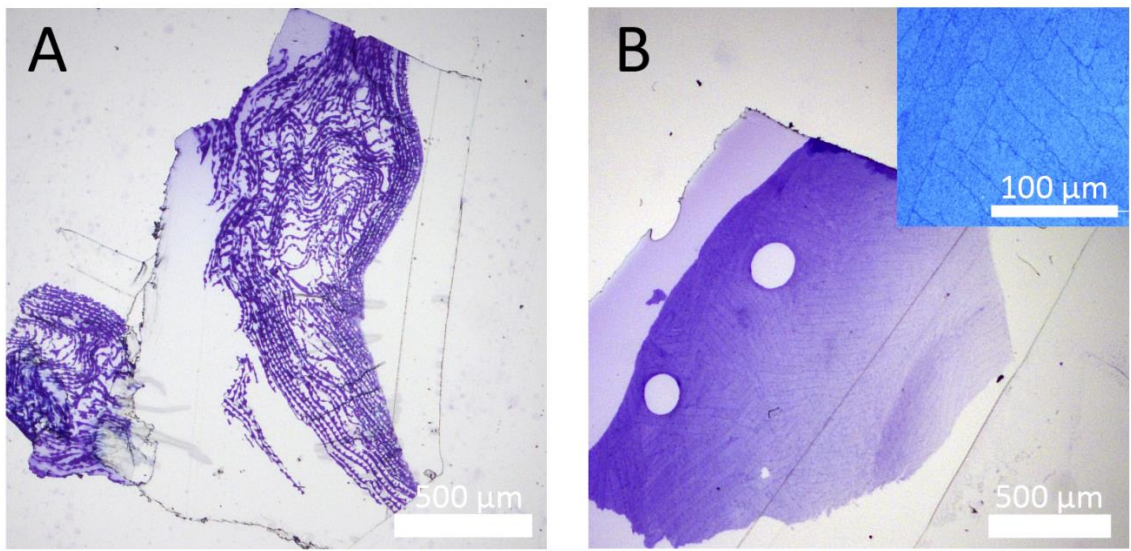

**Figure S3.** Comparison of two processes. Semi-thin sections of a sample after fibrillogenesis following the protocol elaborated (A), and section of a sample freeze-casted and directly dipped into PBS10X (B). Inset: close up where the walls can be observed but the pores are occluded.

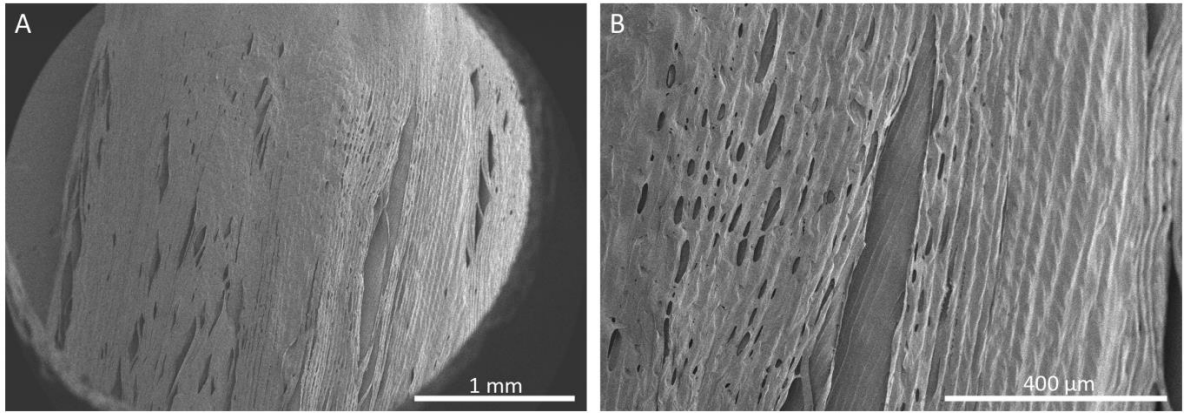

**Figure S4.** SEM pictures of the outside of a lyophilized scaffold after freeze-casting.

**Movie S1. (Not available on ArXiv)** SHG and fluorescence 3D reconstitution from confocal microscopy of collagen scaffold after fibrillogenesis and seeding with NHDFs for three weeks. (A) Reconstitution of an exterior surface of the scaffold. (B) Reconstitution of cell gradient from a side of the scaffold. (C) Reconstitution of NHDFs on parallel walls with ridges inside the scaffold. Green: Actin; red: collagen SHG signal. Visualization window for 3D reconstitutions: 591 x 591 µm.

**Movie S2. (Not available on ArXiv)** SHG and fluorescence 3D reconstitution from confocal microscopy of collagen scaffold after fibrillogenesis and seeding with C2C12 myoblasts for 11 days. (A) Reconstitution of the colonization through open pores at a scaffold extremity. (B) Reconstitution of a myoblasts gradient of colonization on a side of the scaffold. Green: actin; blue: DAPI; red: SHG signal. Visualization window for 3D reconstitution: 738x738 µm.